**Balázs Lajos György: Paál György és a kozmológia forradalma**

(halálának 25-ik évfordulójára)

L.G. Balázs

# G. Paál and the cosmological revolution (for his memory)

abstract

György Paál the Hungarian cosmologist died in 1992. This article was published twenty-five years later in the Hungarian Astronomical Association (HAA, MCSE in Hungarian) 2017 yearbook for his memory. After short introduction of the history of cosmology we briefly introduce the XX. century cosmological discoveries. For his early years G. Paál studied the large-scale inhomogeneity in the Universe. He was one of the first astronomer who realized some periodicity in the quasar redshift distribution. In the early 90s he also studied the galaxy redshift distribution in the so-called pencil-beam survey and suggested the quasi periodicity can be cause of the late inflation in the Universe namely the Lambda cosmological constant.

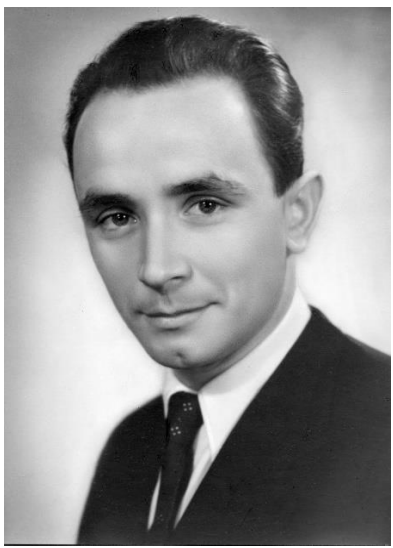

**1. ábra.** Paál György (Paál család arhivum)

Derült éjszakákon a csillagos égbolt látványa ősidők óta gyönyörködteti az embert. Ráboruló sötét boltozat ezüstösen szikrázó csillagokkal. Az ember már korán észrevette, hogy az égbolt alapvetően különbözik földi világunktól. Bármerre ment, nem jutott hozzá közelebb, illetve nem kerül tőle távolabb. A tudomány hosszú utat járt be addig, amíg felismerte, hogy ez a látvány a tér és az idő földi léptékkel mérve úgyszólván határtalan mélységét rejti. Az ember szembesült azzal, hogy földi létének térbeli és időbeli korlátai eltörpülnek a kozmikus térbeli és időbeli távlatok mellett. Azt az érzést, amit ennek a felismerése kelt bennünk találóan fogalmazta meg Blaise Pascal a 17. században élt francia matematikus és fizikus Gondolatok című művében: *„Ha elgondolkozom rajta, milyen rövid ideig tart az előtte volt és utána következő öröklétbe vesző életem, milyen kicsi az a tér, amelyet betöltők, sőt az is, amit látok, az általam nem ismert és rólam nem tudó terek végtelenségében elmerülve, megrémülök, és döbbenten kérdezem, miért vagyok éppen itt és nem másutt, mert ennek nincs semmi magyarázata, miért inkább itt, mint ott, miért éppen most és nem máskor."*[1] Az emberi megismerés alapvető igénye, hogy megpróbáljon ezekre a kérdésekre válaszolni. A kozmológia a tudománynak az a területe, amely a bennünket körülvevő világ térben és időben legnagyobb léptékű szerkezetét próbálja feltárni. A Világ egészének a jellemzése és megértésének az igénye kultúrtörténete során végig kísérte az embert.

**Naiv kozmológiák**

A teljesség igénye nélkül néhány elképzelés az ókorból:

Babilóniai kozmológia: a ránk maradt babilóniai dokumentumok szerint (kb. Kr.e. 3000 körül) a Föld lapos, amely a végtelen „káosz vizén" úszik. A Föld és az Égbolt egységet alkotnak a végtelen „káosz vizén". A Föld lapos, és az Égbolt merev kupolája távol tartja a külső „káosz óceánt".

Az eleai görög iskolát megalapító Parmenidész (született kb. Kr.e. 515 körül) szerint a Világegyetem véges és gömb formájú, amely nem változik, tökéletes és szükségszerűen időtlen állandó, ami azt is jelenti, hogy nem keletkezett és nem is szűnik meg. Egységes egészet alkot, és nem osztható részekre. Űr, tehát olyan hely, ahol semmi sincsen, nem létezik. Sokféleség és változás csak az érzéki megismerés korlátaiból adódik. Parmenidész szerint az időbeli és térbeli korlátok a létező egészhez képest önkényesek és viszonylagosak.

Arisztotelész (Kr.e. 384–322) Univerzuma geocentrikus, állandó, időben változatlan, térben véges kiterjedésű, időben végtelen. A gömb alakú Földet koncentrikus égi gömbök veszik körül. Az Univerzum változatlanul, örökké létezik. A négy klasszikus elemen (föld, víz, levegő tűz) túlmenően a mindenütt jelen levő éter tölti ki.

Arisztarhosz (kb. Kr.e. 280) a Napot helyezte a Világegyetem központjába. A Föld naponta megfordul a tengelye körül és körpályán kering a Nap körül. A csillagok egy Nap középpontú gömbön helyezkednek el.

Ptolemaiosz (Kr.u. 2. század) Arisztotelész nyomán geocentrikus Világot képzelt el, amely a mozdulatlan Föld körül forog. A bolygók körpályán mozognak, a szintén körpályán mozgó középpont a deferrens körül. Ennek a modellnek az volt az előnye, hogy segítségével az akkori mérési pontosságon belül meg lehet jósolni a bolygók égen elfoglalt helyét. A megfigyelésekkel való, a mérési pontosságon belüli egyezésnek köszönhetően ez a modell évszázadokon keresztül megingathatatlannak tűnt.

A középkor általánosan elfogadott világképe szerint a Világegyetem időben véges, és ennek megfelelően van kezdete is. Filoponusz (Kr.u. kb. 490 – 570) alexandriai keresztény filozófus szerint a Világegyetem időben véges, ennek megfelelően volt kezdete, szemben az ókori görögök állításával, akik időben végtelennek gondolták. Logikai érveket hozott fel a véges Világegyetem mellett a muzulmán Alkidus, a zsidó filozófus Saadia Gaoun, illetve a muzulmán teológus Algazel.

---

[1] Blaise Pascal: Gondolatok, ford. Pődör László, Gondolat Kiadó, 1978.

**A távcső felfedezése – a természettudomány forradalma**

A középkori világkép évszázadokig stabilnak hitt talpazatán az első súlyos repedést a 16. század elején Kopernikusz rendszere jelentette, amely a Föld helyett a Napot helyezte a Mindenség középpontjába. A természettudomány fejlődésében az igazi áttörést azonban a távcső 1608-ban történt felfedezése indította el, amely a holland Hans Lippershey (1570 –1619), a nevéhez fűződik. Ő azonban a távcsövet nem akarta tudományos célra használni, alapvetően üzleti szempontok vezérelték. A tudománytörténet a csillagászati alkalmazást Galileo Galileihez (1564-1642) köti. Távcsöves megfigyelései a csillagászatban áttörést jelentettek, de a szabadesés és a tehetetlenség törvényének a felfedezésével előfutára volt a Newton-i mechanikának is.

A tudományos világképben hatalma áttörést hozott Newton tömegvonzás törvényének a felfedezése. E szerint a Földön észlelhető nehézkedésért ugyanaz az erő felelős, amely a bolygókat a Nap körüli pályán tartja. Az „égi" és „földi" világ ugyanazoknak a törvényeknek engedelmeskedik. Ezt az elvet fejezi ki az Isaac Newton (1642–1727) által megalkotott mechanika három axiómája is. Úgy gondolta, hogy az összes jelenség (akár égi, akár földi) az abszolút térben játszódik le. Az abszolút tér maguktól a benne lejátszódó jelenségektől függetlenül létezik. Az időt is ugyanígy a jelenségektől független abszolút létezőnek tekintette.

Newton a Világot időben állandónak és végtelennek képzelte. A tömegvonzás általánosan érvényes, azaz bármely két része kölcsönösen vonzza egymást. Az eloszlása nagy léptéken egyenletes, ennek megfelelően a gravitációs erők kioltják egymást.

A 18. században Immanuel Kant (1724 - 1804) és Johann Lambert (1728 - 1777) a Világegyetemet állandó állapotúnak és végtelennek képzelte el, ahol az anyag különböző léptékeken csomósodik, a csomók újabb nagyobb kiterjedésű csomókká állnak össze meghatározott hierarchiát alkotva a végtelenségig.

A 19. században a fizika rohamos fejlődésnek indult. Megszületett a termodinamika és az elektrodinamika. Úgy tűnt a fizika minden megfigyelhető jelenségre korrekt magyarázatot tud adni. Kialakult a század végére a klasszikus fizika igen impozáns épülete. A kozmológiában azonban lényeges előrelépés nem történt, két látszólag jelentéktelen apróságtól eltekintve. A két látszólag jelentéktelen apróság közül az első az Heinrich Olbers (1758–1840) által 1823-ban megfogalmazott fotometriai, míg a másik a Hugo von Seeliger (1849– 1924) féle gravitációs paradoxon volt 1895-ben.

Olbers azt az egyszerű kérdést tette fel, hogy miért van éjszaka sötét? Amennyiben ugyanis a Newton-féle homogén, végtelen világot fogadjuk el, akkor könnyen kiszámítható, hogy az egész égboltnak úgy kellene fénylenie, mint a Nap. A csillagok fényessége a távolság négyzetével fordított arányban csökken. Vagyis egy csillag 2-szer, 3-szor nagyobb távolságban 4-szer, 9-szer halványabb. Ugyanakkor 2-szer, 3-szr nagyobb távolságban 4-szer, 9-szer több csillag van.

Ennek az az eredménye, hogy összesen a megfigyelőhöz ugyanannyi fény jut. Amennyiben tehát a Világegyetem végtelen, és egyenletesen tele van szórva csillagokkal, akkor nem lenne éjszaka, szemben a legegyszerűbb hétköznapi tapasztalattal,

A Seeliger féle gravitációs paradoxont már nem lehet ilyen egyszerűen szemléletessé tenni. Ennek a paradoxonnak az a lényege, hogy feltételezve a tömegvonzás törvényének általános érvényességét a Newton féle Világegyetemben, ahol nagy léptékben az anyag egyenletesen oszlik el, azaz nincs kitüntetett hely, a gravitációs erőt ugyan a távoli tömegek kölcsönösen kiegyenlítik, de az egész rendszer labilissá válik, és ez a végtelen, egyenletes tömegeloszlás nem maradhatna fenn.

**A kozmológia forradalma a 20. században**

Ugyan a Seeliger paradoxont már csak a 19. század végén fogalmazták meg, de az Olbers paradoxon már a 20-as években ismert volt. Valószínűleg Newton nagy tekintélye miatt világképe általánosan elfogadott maradt.

A fizika viharos fejlődése megteremtette az alapját a műszaki tudományok, és ezen keresztül a technika viharos fejlődésének is. A technika viharos fejlődése lehetővé tette egyre nagyobb teljesítményű (méretű) távcsövek építését is. A 20. század elején, 1917-ben állt üzembe a 2,5 m-es tükrével a maga korában óriásnak számító, Hooker távcső (2. ábra). A kaliforniai Mount Wilson Obszervatóriumban álló távcső 1917 és 1949 között a világon a legnagyobb volt.

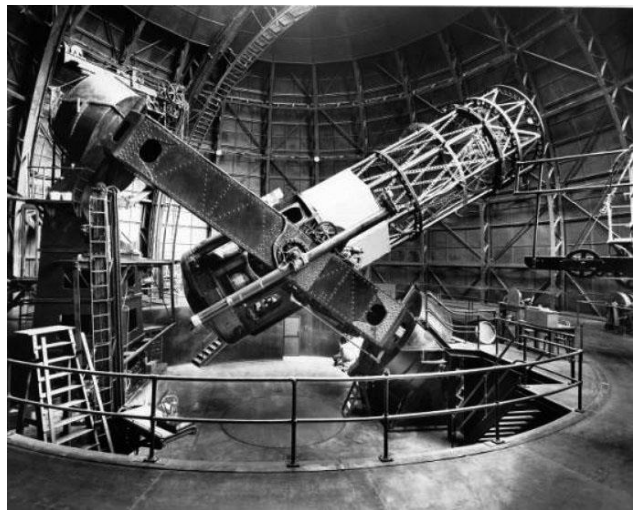

**2. ábra. Az 1917-ben, üzembe ált 2,5 m-es Hooker távcső.** (Courtesy of The Observatories of the Carnegie Institution for Science Collection at the Huntington Library, San Marino, Calif.)

A Hooker távcső üzembe állításának évében látott napvilágot Albert Einstein világmodellje. Ez a modell már az általa a Porosz Tudományos Akadémián 1915. november 25-én bemutatott téregyenlet, illetve az 1916-ban közzétett általános relativitáselmélet felhasználásával jött létre. Einstein, 1917-ben közzétett modellje pozitív görbületű (önmagába visszahajló) térben homogén, illetve időben állandó volt, vagyis az anyag nagy léptéken tekintve állandó sűrűséggel rendelkezett, és nem mozgott. Ilyen megoldást az 1915-ben bejelentett egyenlete nem adott. Azért, hogy kapjon ilyen megoldást, egyenletéhez hozzáadott egy $\Lambda$ kozmológiai állandót.

Einstein téregyenlete a tér geometriai tulajdonságait köti össze az azt létrehozó anyag fizikai sajátságaival (energiájával és impulzusával). Ennek megfelelően az állandónak aszerint tulajdoníthatunk fizikai értelmet, hogy az egyenletben szereplő fizikai, vagy geometriai tagokat egészítjük ki vele. Egyenletének ezzel az állandóval történő kiegészítését Einstein később élete legnagyobb tévedésének tekintette.

Einstein eredeti, 1915-ben közzétett egyenletét használva Alexander Friedmann orosz tudós 1922-ben térben zárt, időben táguló megoldást hozott nyilvánosságra. A megoldás szerint a tér a tágulást követően újra összehúzódik, és ez a folyamat a végtelenségig ismétlődik. Két évvel később, 1924-ben egy hiperbolikus, térben nyílt megoldást közölt, amely korlátlanul tágul a végtelenségig.

Georges Lemaître belga pap és fizikus 1927-ben Einstein egyenletét használva Friedmanntól függetlenül hozzá hasonló eredményre jutott. Sajnos Lemaître munkáját egy helyi belga lapban publikálta, így ez az eredmény nem vált széles körben ismertté, jóllehet Einstein tudott róla. Lemaître világmodellje szupersűrű kezdeti állapotból („ősatomból") érte el jelenlegi formáját, így őt tekinthetjük a Nagy Robbanás (Big Bang) modell „ősatyjának".

A kozmológia elméleti viharos fejlődésével párhuzamosan egy évek óta tartó vita is felforrósodott. Az Egyesült Államok Mount Wilson Obszervatóriumának csillagásza Harlow Shapley amellett kardoskodott, hogy a Világegyetem lényegében a Tejútrendszerből épül fel, és vele szemben állt Herber D. Curtis, aki amellett érvelt, hogy Galaxisunk csupán egy sziget a Világegyetemben, és nagy távolságból ugyanolyan halvány ködfoltnak látszana, mint amilyeneket távcsöveinkkel nagy számban megfigyelünk.

Ez a két nézet egy azóta híressé vált nyilvános vitában csúcsosodott ki, amely 1920. április 26-án zajlott le Washingtonban az USA tudományos akadémiájának ülésén. A vitát végül is Edwin Hubble oldotta fel. A Mount Wilson-i 100 hüvelykes Hooker távcső segítségével az Andromeda-galaxist sikerült csillagokra feloldania, és ezek közül néhány a cefeidákra jellemző fényváltozást mutatott. A Henrietta Lewitt által 1912-ben talált periódus-fényesség relációt alkalmazva megbecsülte az Andromeda galaxis távolságát, amely azt messze a Tejútrendszer határain kívül helyezte (3a ábra). A korszakalkotó felfedezést Hubble az Amerikai Csillagászati Társaság 1925. január 1-én tartott ülésén jelentette be.

Edwin Hubble 1929-ben egy másik korszakalkotó felfedezést jelentett be. Az Andromeda ködéhez hasonlóan meghatározta több galaxis távolságát, és azt találta, hogy a színképükben megfigyelhető vonalak a laboratóriumban mért hullámhosszukhoz képest a távolságukkal arányosan a hosszabb hullámhosszak (a vörös) felé tolódnak el (3b ábra). Ezt az összefüggést a felfedezőről Hubble törvénynek nevezték el, jóllehet létét 1927-ben Lemaître elméletileg már megjósolta.

**A kozmológia Magyarországon – Paál György pályájának kezdete**

Hubble és kortársai mérésinek pontossága mai szemmel nézve sok kívánni valót hagynak maguk után, mégis az 1930.as évektől kezdve az Einstein egyenletek Friedmann és Lemaître által kapott megoldásait a megfigyelt Világegyetemre alkalmazva megszületett a Kozmológiai Elv, ami azt állítja, hogy az Univerzumban nagy léptéken nincs kitüntetett hely, illetve irány, vagyis az homogén és izotróp. Ez az elv hamarosan széles körben elterjedt.

A kozmológiának a 20-as és 30-as években lezajlott forradalma azonban Magyarországot érintetlenül hagyta. A Svábhegyen az állam által Konkoly Thege Miklóstól átvett, de a trianoni béke következtében elcsatolt ógyallai intézet helyett 1921-28-ban épült új létesítmény kutatási témái között nem szerepelt az extragalaxisokkal, illetve a kozmológiával kapcsolatos téma. Nemzetközi szinten az intézet távcső parkja a változó csillagok és a Nap körüli kisbolygók kutatását tette lehetővé. A kozmológia a nagy távcsövekkel rendelkező gazdag országok privilégiuma volt. Sajnos, Magyarország nem tartozott ezek közé.

Az 50-es évek vége felé mégis Magyarországon is elindult valami. Marx György az ELTE TTK fizikus professzora vezetésével, arra alapozva, hogy a csillagok energia termelésének melléktermékeként keletkező neutrínók nagy számban találhatók a Világmindenségben, kutatás indult, hogy ezeknek a részecskéknek a gravitációs hatása befolyásolja-e a nagyléptékű kozmikus struktúrák (pl. galaxis halmazok) stabilitását?[2]

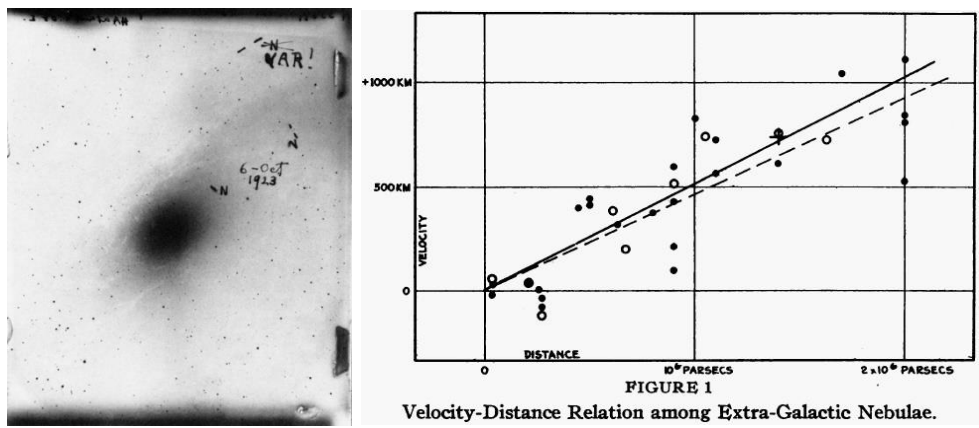

**3. ábra: Hubble felvétele az Andromeda galaxisról (a). Az első Hubble diagram (b).** (Hubble, 1929)

A magyar kutatók rendelkezésére álló csillagászati távcsövek terén lényeges változás jelentett a mátrai Piszkés tetőn az 1958-1961 években épült új megfigyelő állomás. Ezen az állomáson kezdte el működését 1962-ben egy 90 cm tükörátmérőjű Schmidt optikai rendszerű teleszkóp. A távcső csillagászati műszereknél nagynak számító 5°x5° látómezejét felhasználva kutatni kezdték az extragalaxisokban fellángoló szupernóvákat. A szupernóvák kutatásához szükséges volt az akkor már kereskedelmi forgalomban beszerezhető Palomar Atlaszra, amely lehetővé tette annak az eldöntését, hogy s Schmidt teleszkóppal felfedezett újonnan fellángolt forrás halvány formában már korábban is létezett-e, vagy valóban új felfedezés.

---

[2] E sorok szerzője is a kozmikus neutrínók galaxishalmazokra gyakorolt gravitációs hatásából az ő vezetésével írta diplomamunkáját.

Az atlasz a Palomar hegyen 1949-ben felállított 1,8 m tükörátmérőjű Schmidt teleszkóppal készült (4. ábra). 1956. június 20-ig a felvételek 99%-a elkészült. A hiányzó 1%-ot 1958. december 10-ig pótolták. A felmérés során 36 cm (14 hüvelyk) élhosszúságú fotólemezeket használtak, amelyek az égbolt nagyjából 36 négyzetfokos területét fedték le.

Minden égi területet kétszer fényképezetek le: egy vörös, illetve egy kékérzékeny lemezre. A felmérés az északi pólustól a -30º deklinációig az egész égboltot lefedte, sőt egyes területeken a -34º deklinációt is elérte, és 936 lemezpárt tartalmazott. A fényesség határ nagyjából 22 magnitúdó volt. A kereskedelmi forgalomba került fotómásolatoknál ez mintegy 21 magnitúdóra csökkent. A vörös érzékeny lemezeknél ez valamivel rosszabb, a kékeknél valamivel jobb olt. A svábhegyi csillagászati intézet (akkori nevén az MTA Csillagvizsgáló Intézete) fiatal munkatársa, Paál György felismerte, hogy az intézet által beszerzett Palomar Atlasz egyedülálló lehetőséget biztosít a magyar kutatók számára is, hogy vizsgálataikat a Világegyetem távoli térségeire is kiterjesszék.

Paál György 1934. december 31-én született Budapesten. Általános iskolai tanulmányait : a XI. kerületi Bocskai úti népiskolában színjeles, azaz kitűnő eredménnyel végezte. Az Apáczai Csere János gyakorló általános gimnázium reál tagozatán szerzett érettségi bizonyítványt 1953. június 12-én, ugyancsak kitűnő eredménnyel. Az 1953/54 tanévtől 1956/57 tanévig terjedő időben az Eötvös Loránd Tudományi Egyetem, Természettudományi Kar matematika-fizika szakán tanult. Az államvizsgát lebonyolító bizottság 1957. évi július hó 23. -i határozata alapján okleveles középiskolai tanár végzettséget kapott. Az államvizsgát is kitűnő eredménnyel tette le.

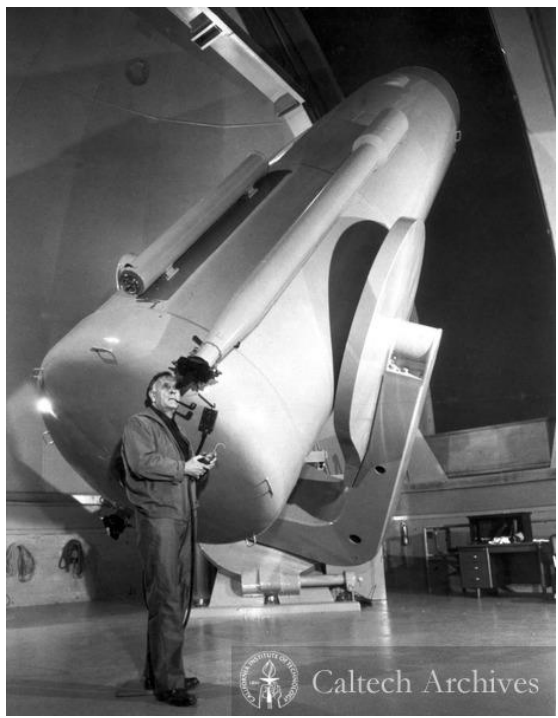

**4. ábra: Edwin Hubble és a Palomar hegyi Schmidt távcső.** (Caltech Archives)

Az 1956-os forradalom után keletkezett menekült hullámban sok tehetséges fiatal hagyta el hazánkat. Ez alól a svábhegyi csillagászati intézet sem volt kivétel. Detre László igazgató a nyugatra távozott fiatalokat frissen diplomázottakkal igyekezett pótolni. Így került 1958-ban Paál György is az intézetbe. Abban az időben az intézet már nemzetközi hírnévre tett szert az RR Lyrae típusú változócsillagok kutatásában. Paál György is ezeknek a csillagoknak az észlelésébe kapcsolódott be. Hamarosan kitűnt kivételes munkabírásával és lelkiismeretességével.

Az észlelőmunka mellett rendszeresen képezte magát. Ebben nagy segítséget nyújtott az intézet nemzetközi összehasonlításban is kiváló könyvtára. A mindenkori vezetők, és jelesen az abban az időben hivatalban levő Detre László igazgató is kiemelt fontosságot tulajdonított a könyvtár fejlesztésének. Minden fontosabb csillagászati folyóirat (Astrophysical Journal, Astronomical Journal, Monthly Notices of the Royal Astronomical Society, Astronomische Nachrichten, stb.) a kezdetektől indulva hiánytalanul rendelkezésre állt. A két világháború okozta kiesést rövid időn belül pótolták.

Olvasmányai során Paál György figyelme fokozatosan a Világegyetem nagyléptékű szerkezete felé irányult, és elhatározta, hogy ezzel a témával fog foglalkozni. A témaváltás első dokumentuma egy a Nature c. folyóirathoz beküldött olvasói levél volt 1961-ben. A levelet Balázs Bélával közösen írták, aki frissen végzett matematikusként szintén az 1956-os forradalom után nyugatra távozott kutatók helyére került az intézetbe. A levélben ahhoz a kérdéshez szóltak hozzá, hogy a kozmológiában az adott mérések és hibák szintjén mennyire lehet különbséget tenni az egyes modellek között. Hangsúlyozták, hogy a modellekre mindenképpen szükség van, de csak olyanokra, amelyek között az estleges jövőbeli mérésekkel különbséget lehet tenni, amely azonban nem jelent okvetlenül igazolást is.

**Paál György kutatómunkájának főbb állomásai**

Tudományos munkásságát nem sokkal 1992. március 22-én bekövetkezett halála előtt 5 téma vizsgálata köré csoportosította:

A. A korai években (1964-1971) gondolatokat fogalmazott meg a Világegyetemben észlelhető nagyléptékű inhomogenitásokról.
B. Ezeket az inhomogenitásokat a kövezőkben bizonyos fejlődési effektusok lehetséges eredményének tartotta
C. Az első érveket és elemzéseket, amelyekben a kazárok vöröseltolódásában megfigyelhető periodicitást a kozmikus tér globális szerkezetével magyarázhatók, 1970-19971-ben fogalmazta meg.
D. Később a galaxisok vöröseltolódásában megfigyelhető periodicitást kapcsolatba hozta a kozmikus tér globális szerkezetével, illetve a Világegyetem expanziójának a gyorsulásával.
E. Vizsgálta a Világegyetem korai inflációjának lehetséges változatait.

A továbbiakban Paál György munkásságát az általa javasolt beosztás szerint próbálom áttekinteni. Az „E" pontban összefoglalt munka időben korábban történt, mint a „D". A tárgyalásnál én az időbeli sorrendet követtem.

*A korai évek, a Világegyetemben észlelhető nagyléptékű inhomogenitások vizsgálata*
*(1964-1971)*

A 20. százat 50-es éveire már számos vizsgálat történt a Világegyetem nagyléptékű anyageloszlásának a tanulmányozására. Ennek során különböző irányokban és határfényességekig analizálták a galaxisok előfordulási gyakoriságát. Természetesen, a galaxisok távolsága kapcsolatban van látszó fényességükkel, de az abszolút fényességükben meglevő nagy különbségek erősen korlátozzák az ilyen természetű vizsgálatok statisztikai hatékonyságát. Ezzel szemben a galaxisokból felépülő halmazok kiválóan alkalmasak az anyag nagyléptékű eloszlásának a vizsgálatára

A térbeli eloszlás vizsgálatához a halmazok esetében is szükségünk van valamilyen a távolsággal kapcsolatban levő adatra. Ilyen kézenfekvő adat a halmaz látszó átmérője, illetve tagjainak a fényessége. Egyszerű matematikai statisztikai tény, hogyha a halmazt alkotó galaxisok abszolút fényességét nagyság szerint sorba rendezzük, akkor a 2-ik, 3-ik, stb. legfényesebb elem fényességének a szórása a sorba rakott elemekénél sokkal kisebb. Megfigyelési tény, hogy galaxisoknál a 10-ik legfényesebb elem abszolút fényességének a szórása mindössze m=0.35 magnitúdó. Az persze probléma, hogy a halmazok az egyedi galaxisoknál sokkal kisebb számban találhatók.

A Palomar-hegyi Schmidt távcsővel az ötvenes évek második felében készített égbolt felmérés az égbolt lefedettségében, valamint kozmikus távolsági mélységében az addigi legátfogóbb volt. Ezt a felmérést felhasználva, amely alapján a Magyarországon is beszerzett Palomar Atlasz is készült, Georg O. Abell, a Mount Wilson Obszervatórium munkatársa több tízezer galaxis halmazt és csoportot azonosított. Ezek közül közel kétezer volt elegendően gazdag ahhoz, hogy homogén mintaként statisztikus vizsgálatra alkalmas legyen. A további statisztikus vizsgálatra alkalmas homogén minta 1682 halmazt tartalmazott.. A felmérés az égbolt északi pólustól a -27° deklinációig terjedő részét ölelte fel. Ez a nagyszabású munka felkeltette Paál György figyelmét is, és az eredményeket kritikai vizsgálat alá vette.

Abell galaxishalmaz katalógusának egyik fontos célja volt, hogy segítségével ellenőrizni lehessen a kozmológiai modellek érvényességét. Elfogadva egy modellt valamely objektum színképében mért vöröseltolódás értékéből meghatározhatjuk a Hubble törvény alapján a távolságát. Az eredmény természetesen függ az alkalmazott modelltől. Amennyiben azonban a távolságot ettől független úton is meg tudjuk határozni, akkor van esélyünk a modell érvényességének az ellenőrzésére is.

Fentebb már említettem, hogy a halmaz 10-ik legfényesebb elemének az abszolút fényesség m=0.35 magnitúdó pontossággal megbecsülhető, feltéve, ha a közeli és távoli halmazokon belül az abszolút fényesség eloszlása, az ún. luminozitási függvény mindkét halmaz esetén ugyanaz. Ezt természetesen nem lehet garantálni. Abell ezt a nehézséget úgy próbálta kiküszöbölni, hogy bevezette az ún. „gazdagság" osztályokat (richness class). A halmazokat a bennük azonosítható galaxisok száma szerint sorolta. Összesen 5 osztályt állított fel a „legszegényebb" csoportba azok tartoztak, amelyeknek 30-79, míg a leggazdagabbakba azok, amelyekben több mint 300 halmaztag volt. (Az egyik leggazdagabb Abell halmaz, az 1689-es látható az 5. ábrán[3])

A kozmológia elméletek különféle kapcsolatokat állítanak fel a galaxis-halmazokhoz rendelhető látszó fényesség, halmaz átmérő, halmaz tagok száma, valamint a vöröseltolódás között. Paál György hangsúlyozta, hogy ezeket a mennyiségeket az adott modellek ellenőrzésére csak akkor szabad alkalmazni, ha valóban azt mérik, amire vonatkoznak, tehát szisztematikus hibáktól mentesek. Kimutatta, hogy az Abell katalógusban alkalmazott halmaz-átmérő meghatározás a távoli objektumoknál szisztematikus hibát eredményes. Ennek az a következménye, hogy az általa használt eljárás a ténylegesnél kisebb számú halmazt azonosított.

Paál György javaslatot tett a hiba kijavítására. Furcsa módon a halazok korrigált térbeli eloszlása rosszabbul illeszkedett az Einstein egyenletekből kapott térbeli homogenitást és izotrópiát feltételező megoldásokra. Ebből arra következtetett, hogy abban a térrészben, amelyet a mi galaxisunk körül az Abell katalógus átfog (ez nagyjából 700 Mpc sugarú gömb), a halmazok térbeli eloszlása és/vagy mozgása nem követi a modellek által jósolt egyenletes sűrűséget és tágulást[4]. Kimutatta, hogy a megfigyelt eloszlásokat a homogén izotróp modellekkel csak akkor lehet helyesen értelmezni, ha jelentős pozitív görbületet tételezünk fel a megfigyelt térrészben, vagy a Világegyetem tágulása igen gyors ütemben lassul.

A halmazok átmérőjének a meghatározására használt Abell-féle módszer helyet egy szisztematikus hibáktól mentes új módszert javasolt. A módszert a svábhegyi intézet tulajdonában levő Palomar Atlasz felhasználásával személyesen is leellenőrizte. A halmaz középpontjából kiindulva egyre növekvő sugarú körökben leszámolta a galaxisokat, majd az így kapott számértékeket a körök sugarának függvényében ábrázolta. A kapott görbéken meghatározta azt a pontot, ahol annak menete az egyenestől a legjobban eltért. Ennek a pontnak a középponttól való távolságát fogadta el a halmaz sugarának (5. ábra).

---

[3] Az Abell 1689 galaxis halmaz 2,2 milliárd fényév (677 Mpc) távolságban található a Virgo csillagképben. Ez az egyik legnagyobb tömegű ismert galaxis halmaz, amely ennek következtében jelentős gravitációs lencse hatást fejt ki, torzítva a látó irányban mögötte levő galaxisok képét.

[4] J. Richard Gott III és Mario Jurič, a Princeton-I Egyetem kutatói a Sloan Digital Sky Survey (SDSS) méréseire alapozva 2003-ban bejelentették a Nagy Sloan Fal (Great Sloan Wall) felfedezését, amely 400 Mpc méretű, galaxisokból álló szuperhalmaz, tőlünk 300 Mpc távolságban. 2014-ben Brent Tully, a Hawaii Egyetem kutatója, és munkatársai a galaxisoknak az univerzális Hubble törvényétől eltérő mozgásában egy 1043 milliárd fényév átmérőjű szuperhalmazt találtak, amelynek a Tejútrendszer is része. A halmazt, Laniakea"-nak nevezték, amely Hawaii nyelven a mérhetetlen égboltot jelenti.

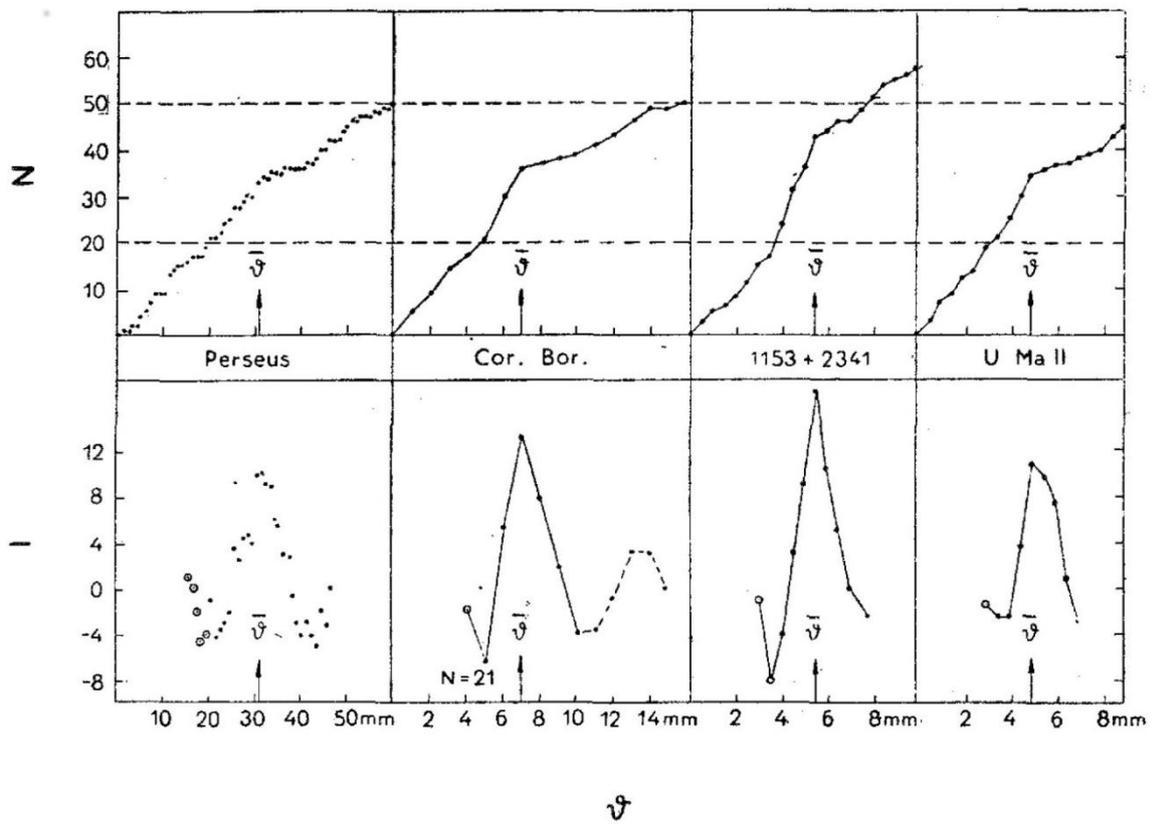

**5. ábra: A Paál György által javasolt eljárás a galaxishalmazok méretének becslésére. A felső sor (N) a halmaz középpontjától kiindulva a galaxisok teljes számát, az alsó sor (I) ennek megváltozását mutatja.** (Paál, 1965)

Az új mérések is megerősítették azt a korábban kapott eredményét, hogy a tőlünk mért nagyjából 400 Mpc távolságon belül a tér jelentősen eltér az euklideszitől, vagy pedig ebben a térrészben az anyag sűrűségében és/vagy mozgásában jelentősen eltér a homogén izotróp esettől. Megvizsgálta, de elvetette azt a lehetőséget, hogy a megfigyelt effektus annak lenne a következménye, hogy a mért vöröseltolódást nem a távolodásból adódó Doppler effektus okozza.

*Az anyag nagyléptékű eloszolásában megfigyelhető inhomogenitások eredete és fejlődése*

A kozmikus anyag nagyléptékű inhomogenitásainak vizsgálatánál alapvető kérdés az, hogy ezek mennyire őrizték meg a születésükkor fennálló ún. kezdeti feltételeket. Azaz, jelenleg megfigyelhető állapotukból ezek a körülmények mennyire olvashatók ki.

Fentebb már említettem, hogy a kozmológiában a múlt század 20-as és 30-as éveiben lezajlott forradalmi átalakulás nyomán a Világegyetem nagyléptékű homogenitása és izotrópiája széles. körben elfogadottá vált. Időben visszafelé követve a homogén izotróp eloszlás tágulását, természetesen adódott a kezdeti igen sűrű ősállapot létének feltételezése, amelyből a jelenlegi struktúrákban igen gazdag állapot létrejött.

A fény véges terjedési sebessége miatt a kozmológiailag nagyobb távolságban levő objektumokat időben korábbi állapotukban látjuk. A jelenleg ismert legtávolabbi objektumok esetében ennek a „visszapillantási időnek" a hossza a távcsöveinkkel megfigyelhető Világegyetem korával összemérhető. Az eltérő távolságban levő objektumok szerkezetének összehasonlításával nyomon tudjuk követni a különféle kozmikus struktúrák (pl galaxishalmazok) kialakulását és fejlődését.

Az eredetileg „sima" kozmikus anyag feldarabolásába és formálásában n a gravitáció játszotta a főszerepet. A sűrűsödéseket/ritkulásokat követő gravitációs zavar hatására lefűződő anyagfelhők már nem követik a kozmikus anyag nagyléptékű általános tágulását, hanem önálló életet kezdenek. Alapvető kérdés, hogy a leszakadt felhő gravitációs ereje elegendő-e ehhez az önálló élethez, vagyis gravitációsan stabil-e? Az önálló életre kelt felhő tovább aprózódhat, és újabb alrendszerek formálódhatnak ki benne. Ennek a folyamatnak jellegzetes állomását képezik a galaxishalmazok.

A kozmológia forradalmának részekét vetődött fel a kérdés már a múlt század harmincas éveiben, hogy a galaxishalmazokban a halmaz tagja között működő gravitációs erő elegendően nagy-e ahhoz, hogy megakadályozza a halmaz széthullását, azaz ezek időben mennyire stabil képződmények? Fritz Zwicky, a California Institute of Technology munkatársa, a Coma csillagképben levő igen gazdag galaxishalmaz (6. ábra) vizsgálata során 1937-ben arra a következtetésre jutott, hogy a halmazt alkotó galaxisok gravitációs ereje nem elegendő ahhoz, hogy a halmazt alkotó galaxisokat a megfigyelt mozgásukkal szemben együtt tartsa. Tekintve, hogy a halmaz a benne levő galaxisok eloszlását tekintve egyensúlyban levőnek tűnik, léteznie kell még a halmazban olyan tömegnek, aminek a vonzása kiegészíti a távcsöveinkkel látható objektumokét, és csak gravitációs hatása révén érzékelhető. Ez az ún. „sötét anyag".

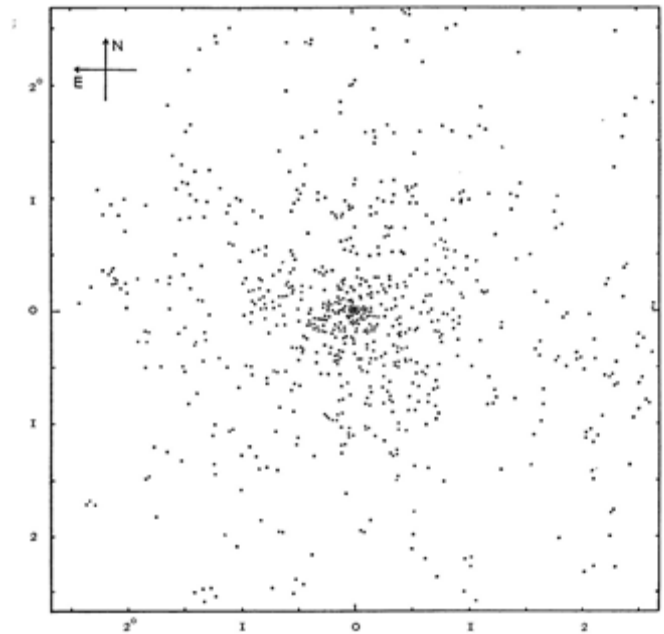

**6. ábra: A galaxisok eloszlása a Coma halmazban. Térbeli eloszlásuk egyensúlyra utal, jóllehet látható tömegük ehhez nem elegendő. Az egyensúly biztosítására feltételezte Fritz Zwicky 1937-ben a sötét anyagot.** (Zwicky, 1937)

Felvetődött a kérdés, hogy ez a csak gravitációsan érzékelhető anyag hogyan népesíti be a halmazt? A valóságot valahol két véglet között lehetett keresni. Az egyik véglet az, hogy a sötét anyag nem tapad a galaxisokhoz, a halmazt nagyjából egyenletesen tölti ki. A másik véglet pedig az, hogy a sötét anyag a galaxisokhoz, és jelesen a zöme a nagyobb tömegűekhez tapad.

Miután Paál György az Abell féle galaxishalmazok vizsgálatával arra a következtetésre jutott, hogy azok térbeli eloszlása/mozgása nem követi a Kozmológiai Elv szabta szabályokat, a70-es években részletesen vizsgálta a kérdést, hogy az észlelt eltérésekben a sötét anyag jelenléte milyen szerepet játszott. Megállapította, hogy a „Nagy Robbanás" után, amikor a Világegyetem anyagának a struktúrákban történő szerveződése elkezdődött valamilyen nagyon hatékony folyamatnak kellett működnie, amely a rendelkezésre álló 13 milliárd év alatt a megfigyelhető galaxishalmazok jelenleg megfigyelhető arcát kialakította . (7. ábra).

Arra a következtetésre jutott, hogy jelentős mennyiségű sötét anyag jelenléte nélkül az a gravitációs hatás, amely a halmazokban a távcsöveinkbe jutó fényt kibocsátó anyagi objektumokból származik, nem elegendő ahhoz, hogy a kezdeti állapotból a rendelkezésre álló idő alatt a jelenlegi arculatukat létrehozza. Az elegendően gyors fejlődéshez a sötét jelenléte mindenképpen szükséges.

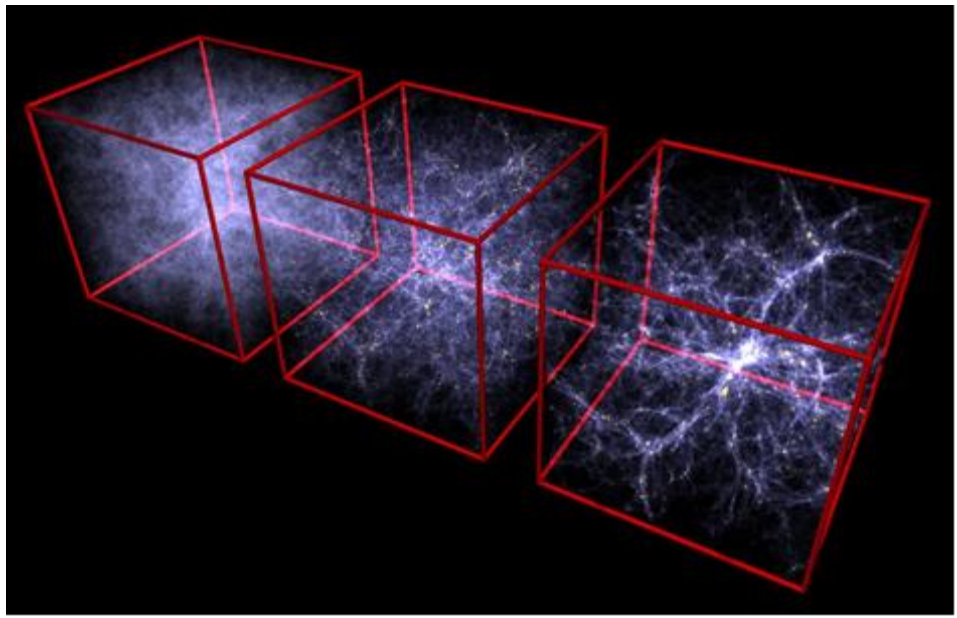

**7. ábra: A kozmikus anyag nagyléptékű strukturálódásának folyamata. Az idő múlásával a szálas szerkezet egyre markánsabbá válik. A gazdag galaxishalmazok a háló csomópontjaiban jönnek létre. A struktúrák kialakulásában a sötét anyag döntő szerepet játszik.** (Credit: Volker Springel/Max-Planck-Institute for Astrophysics)

Ennek a ténynek a felismerése az ő szóhasználata szerint a sötét anyag „újrafelfedezését" jelentette. Rámutatott arra, hogy a sötét anyag hatása rendkívül érzékeny arra, hogy az a halmazban hogyan oszlik el. Konkrét számszerű példákkal igazolta, hogyha a sötét anyag túlnyomó része a halmaz nagytömegű galaxisaihoz tapad, akkor a halmaz fejlődési üteme a sötét anyag nélküli állapothoz képest akár 3 nagyságrenddel is gyorsabb is lehet.

*Első érvek és elemzések a kozmikus tér globális szerkezetéről*

A kvazárok[5] felfedezése 1963-ban mindenképpen új utat nyitott a Világegyetem nagyléptékű szerkezetének a vizsgálatában. Már az első spektroszkópiai mérések során kiderült, hogy a hasonló látszó fényességű nagy galaxisokénál sokkal nagyobb vöröseltolódással rendelkeznek. Amennyiben a vöröseltolódást a Hubble törvény alapján távolság-paraméterként értelmezzük, akkor ez a kazárok számára a megfigyelt galaxisokénál sokkal nagyobb távolságot is jelent. Azoknál a kvazároknál, amelyek gazdag galaxishalmazok irányában fordultak elő sikerült olyan eseteket találni, amikor a vöröseltolódásuk a halmazt alkotó galaxisokéval egyeztek meg. Ennek alapján valószínűsíthető volt, hogy vöröseltolódásuk valóban kozmológiai távolságot jelent.

A 60-as évek végére az ismert vöröseltolódású kvazárok száma jócskán meghaladta a százat. 150 kvazár vöröseltolódásának elemzése arra mutatott, hogy azok nagysága meghatározott értékek köré csoportosul. mintha az valahogyan „kvantálva" lenne. Ez a tény egyes kutatókban kétséget ébresztett a vöröseltolódás kozmológiai eredete iránt. Arra gondoltak hogy ez a kvazárok valamilyen „belső" tulajdonsága, és semmi köze a Hubble effektushoz.

Paál György egy 1971-ben közzétett tanulmányában rámutatott arra, hogy a kvazárok vöröseltolódásának eloszlásában talált szabályos csúcsok nincsenek ellentétben kozmológiai távolságukkal.

A tér és az idő a megfigyelhető anyagi jelenségek legáltalánosabb jellemzői. Bármilyen, a fizika által tanulmányozott jelenség valahol és valamikor történik. Megfelelő vonatkoztatási rendszer[6], illetve időszámítás bevezetésével a megfigyelt jelenség helyéhez és idejéhez számadatokat, koordinátákat, illetve időpontot, is rendelhetünk. Ilyen módon a tér pontjai, és az idő pillanatai matematikailag kezelhető sokasággá válnak. E két sokaság együttesét tér-időnek nevezzük.

A fizikai tér alapvető tulajdonsága az, hogy összefüggő. Ez alatt azt kell érteni, hogy bármelyik pontjából eljuthatunk egy másikba folytonos vonallal úgy, hogy a vonal minden pontja a tér egy pontja. A tér egyszeresen összefüggő, ha a pontjait összekötő bármely folytonos zárt görbe fokozatos zsugorítással egy pontba húzható össze anélkül, hogy eközben belőle kilépnénk. Amennyiben ez nem teljesül, többszörösen összefüggő térről beszélünk.

Ezeket a fogalmakat könnyen szemléletessé tehetjük egy papírlap segítségével. A papíron bármely folytonos zárt görbe fokozatosan ponttá zsugorítható úgy, hogy nem kell a papírlapból kilépnünk. Ragasszuk össze a lap két átellenes oldalát egy hengert formálva. Az így kapott

---

[5] Az első kvazárt (quasar – quasi-stellar adio source), a 3C273 jelű rádióforrás optikai párjaként Maarten Schmidt, a California Institute of Technology (Caltech), munkatársa azonosította 1963-ban a Palomar hegyi 5-m tükörátmérőjű teleszkóp segítségével.

[6] A csillagászatban használt vonatkoztatási rendszer kiinduló pontja a megfigyelő, és bármely jelenséget az égbolton látható helyzetet megadó két szöggel, illetve a látóirány menti távolsággal jellemzünk.

hengeren tudunk olyan folytonos vonalat rajzolni, amit már nem lehet a fenti módon ponttá zsugorítani.[7]

Egy ilyen hengerfelületen élő 2 dimenziós lény saját hátát is láthatja, mivel a hátrafelé elindított fénysugár a henger felületén haladva, azt megkerülve szemből érkezik meg. Valami hasonló történik, mint egy szobában, amelynek falait tükrökkel fedték be (8a ábra). Amennyiben a henger átmérője akkora, hogy a megkerüléséhez szükséges idő a 2 dimenziós lény életkorával is összemérhető, az illető akár saját csecsemő korabeli hátát is láthatja.

Paál György hangsúlyozta, hogy az Einstein egyenleteknek lehetnek olyan megoldásai, amelyekben a megfigyelt periodicitás egyszerűen a térszerkezetének a következménye. Amennyiben a tér, amiben a kozmológiai események lejátszódnak többszörösen összefüggő, akkor Einstein egyenletei nem zárják ki egy olyan megoldást sem, amiben egy objektum a fenti 2 dimenziós esethez hasonlóan többször is látható, de persze a tükörszobához hasonlóan a megfigyelőhöz képest különböző távolságokban és a futási időnek megfelelő korábbi fejlődési állapotában.

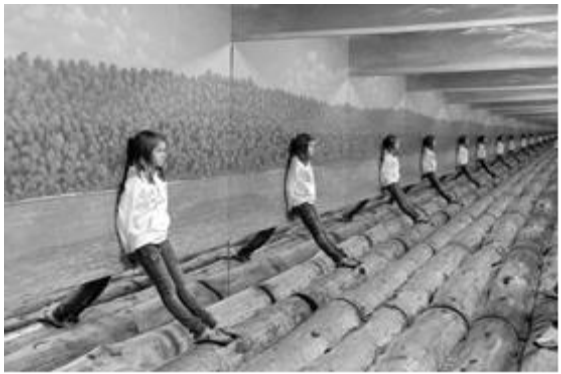 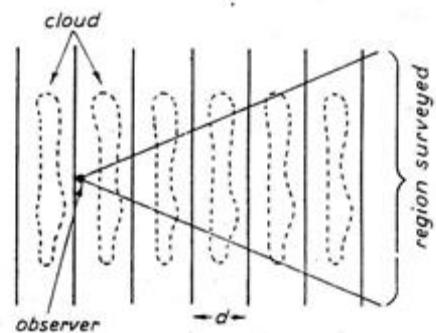

**8. ábra: Egy tükörszobában álló alak megsokszorozva látszik (a). Hasonlóan magyarázta Paál György a kvazárok térbeli eloszlásában észlelt periodicitást (b).** (Paál, 1971)

A kvazárok vöröseltolódásában látható periodicitás tehát ebben az esetben egyszerűen a tér globális szerkezetének a következménye. Ehhez csak azt kell feltételeznünk, hogy van a térben egy „kvazár felhő", amit a tér szerkezetéből adódóan többször is látunk (8b ábra).

A fenti gondolatmenetnek van azonban sebezhető pontja. A periodicitást a vöröseltolódásban látjuk, míg a fenti modell teljesülése esetén a periodicitásnak a távolságokban kellene jelentkezniük. Az eredeti Hubble törvény ugyan azt mondja, hogy a galaxisok színképében a vonalak eltolódásának mértéke arányos a távolsággal, de ez csak kis vöröseltolódás esetén igaz. nagyobb távolságoknál a két mennyiség kapcsolta a megfigyelések leírására használt modell függvénye. A vöröseltolódásban észlelt periodicitás csak speciálisan választott modellek esetén áll fenn. A probléma tisztázásához további megfigyelésekre volt szükség, amelyek a 80-as évek második felében születtek meg.

---

[7] Húzzunk pl. egy vonalat, amely párhuzamos a két nem összeragasztott oldallal. Hengerré formálás után ez a vonal önmagába záródik, de nem húzható össze folytonos zsugorítással egy, a henger palástján levő pontba.

*A Világegyetem korai inflációja*

Paál György 35 évig volt a svábhegyi csillagászati intézet munkatársa. A jelen írásnak nem feladata elemezni, hogy ez alatt az idő alatt miért nem alakult ki rendszeres munkakapcsolat egyik kollégájával sem, beleértve magát a cikk szerzőjét is. Kutatásait tekintve az intézeten belül valahogy végig „magányos farkas" volt. A szomszédos Központi Fizikai Kutatóintézet (KFKI) néhány kutatójával[8] azonban a nyolcvanas évek kezdetétől 1992-ben bekövetkezett haláláig igen eredményes együttműködés alakult ki.

A nyolcvanas évek elején, 1981-ben tette közzé Alain Guth, a Massachusetts Institute of Technology (MIT) kutatója, inflációs kozmológiai elméletét. Az inflációs kozmológiai modell bevezetésére azért volt szükség, mivel a széles körben elfogadott, Einstein téregyenletéből levezetett homogén, izotróp modellekkel kapcsolatban problémák merültek fel.

Guth két súlyos problémát említ: A megfigyelésekkel összehasonlítható modell meghatározásához bizonyos kezdeti feltételeket kell megadni: (1) a táguló Világegyetem korai szakaszában teljes mértékben homogén és izotróp volt, dacára annak, hogy bizonyos tartományai között nem volt oksági kapcsolat (ez a horizont probléma). Továbbá, (2) a Hubble állandó kezdeti értékét igen pontosan be kell állítani ahhoz, hogy napjainkra a tér sem nem negatív, sem nem pozitív görbületű, azaz sík Euklideszi legyen

Guth szerint ezek a problémák eltűnnek, ha feltételezzük, hogy a Világegyetem igen korai szakaszában a szupersűrű anyag állapotában beállt hirtelen változás következtében igen nagy ütemben tágulni kezd („inflálódik" – ahonnan a nevét a modell is kapta). Ilyen átalakulás ebben az állapotban az elemirészek nagy egyesített elméletében (GUT) természetes módon létrejön. Az elképzelés szerint a Világegyetem $10^{-38}$ másodperccel a Big Bang után $10^{29}$ K hőmérsékletre hűl le, és ezen a hőmérsékleten elválik a GUT elmélet szerint addig együtt futó erős, illetve gyenge magerő, valamint az elektromágneses kölcsönhatás. Ez az átalakulás mintegy $10^{-36}$ másodperc alatt a tér mintegy $10^{50}$-szörös „felfúvódását" okozza (9. ábra).

A mikrovilágban mindig jelenlevő kvantum-fluktuációk a gyors, hatalmas méretű tágulás eredményeképpen akkorára növekedhettek, hogy belőlük kiformálódhattak a későbbi kozmikus struktúrák.

---

[8] Diósi Lajos, Holba Ágnes, Horváth István, Keszthelyi Bettina, Lukács Béla, KFKI, Részecske és Magfizikai Kutatóintézet

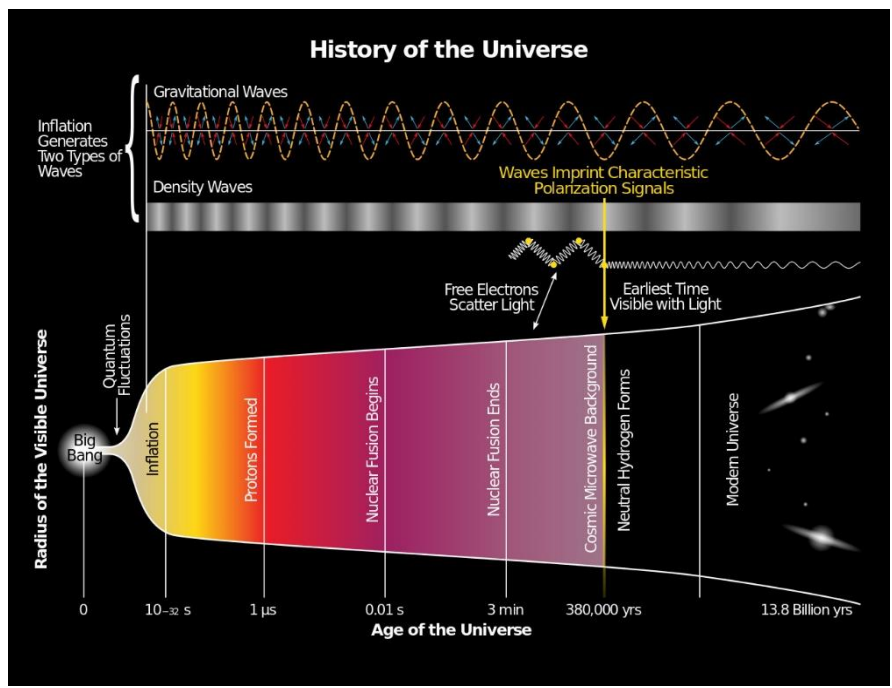

**9. ábra: A Világegyetem inflációs (felfúvódó) modellje. A közvetlenül a Big Bang után bekövetkező kvantum fluktuációt követően kezdetben igen gyors tágulással alakulnak ki a jelenleg megfigyelhető kozmikus struktúrák.** (Credit: bicepkeck.org)

Sajnos ennek az elméletnek is vannak sebezhető pontjai. Ezeken a sebezhető pontokon kívánt javítani a Paál György és Lukács Béla által 1988-ban közzétett megoldás. A szerzők az inflációs elmélet következő problémáit sorolták fel: (1) A kezdeti kvantum-fluktuációknak igen speciális feltételeknek kell megfelelniük, hogy a jelenleg megfigyelhető kozmikus struktúrák létrejöjjenek. (2) A korábbi elméletben fontos szerepet játszó globális homogenitás és izotrópia elvész. Ez csak a megfigyelő által belátható részre igaz, jóllehet számára a világ szükségképpen Euklideszi, amit viszont a korábbi elmélet nem tudott megmagyarázni. (3) közel kritikus anyag sűrűséget kell feltételezni, ami biztosítja, hogy a megfigyelt tágulási sebesség mellett azok a kozmikus struktúrák jöjjenek létre, amelyeket ma megfigyelünk.

A szerzők olyan megoldást javasoltak, ami bizonyos értelemben „arany középút" az eredeti, az Einstein egyenletből levezetett homogén izotróp, illetve a Guth-féle inflációs kozmológiai modell között. A javasolt megoldás az eredeti, Guth-féle igen drasztikus inflációt próbálta „megszelídíteni". Elképzelésüket „langyos inflációnak" nevezték el.

Számításokkal igazolták, hogy a „lágyabb infláció esetén lehetséges, hogy az ősi sűrű anyagban folyamatosan, párhuzamosan keletkezzenek inflálódó helyek, és ebben az esetben már az így kapott megoldás sok tekintetben megőrzi az eredeti homogén izotróp megoldások tulajdonságait, és a kozmológiai elv a Világegyetem belátható részén túl is igaz lehet.

*A Világegyetem gyorsuló tágulása*

A kvazárok vöröseltolódásában megfigyelhető ciklikusság levont következtetések megerősítéséhez további megfigyelési adatokra volt szükség. Ezek az adatok a 80-as évek második felében születtek meg. Így a probléma csaknem 20 éven át Csipke Rózsika-álmát aludta. A megfigyelések az északi és a déli galaktikus pólus irányában két vékony nyalábban történtek, és 5 milliárd fényév (1400 megaparszek) távolságig terjedtek.

A megfigyelések alapján Thomas Broadhurst, a Durhami Egyetem munkatársa, együttműködve Richard Ellissel (Durhami Egyetem), David Kooval (Lick Obszervatórium), valamint Szalay Sándorral (Johns Hopkins Egyetem,, Eötvös Loránd Tudományegyetem), arra az eredményre jutottak, hogy a vizsgált galaxisok eloszlásában 127 Mpc távolságokként csúcsok jelentkeznek. Arra következtetésre jutottak, hogy ezt a periodicitást a galaxisok kialakulására ismert akkor elméletek alapján nem lehet megmagyarázni.

Az effektus magyarázatához Paál György és munkatársai (Horváth István, Lukács Béla, később Holba Ágnes) szerint nem volt szükség új, nagy léptékű struktúrák kialakulásáért felelős mechanizmusra. Egyszerűen vissza kellett térni Paál Györgynek a kvazárok eloszlásában megfigyelhető periodicitás magyarázatára mintegy 20 évvel korábban tett javaslatához, nevezetesen ahhoz, hogy az észlelt periodicitás egyszerűen a tér többszörösen összefüggő természetének a következménye, és nincs szükség semmilyen új struktúra formáló mechanizmus feltételezésére.

Fentebb a tér többszörösen összefüggő voltát egy sík papírlap hengerré formálásával szemléltettem. Ennek előállításához a végtelen síkból ki kellett vágnunk egy véges méretű darabot, és annak két ellentétes oldalát összeragasztva hengert formálnunk. A henger felületéhez rögzített 2 dimenziós világban a fény visszatérhet kiinduló pontjába, és ez akárhányszor előfordulhat. Az ilyen világban élő 2 dimenziós lény tehát ezt e véges méretű hengert végtelennek látja.

Paál György szerint többszörösen összefüggő tér esetén ez a mi 3 dimenziós világunkban is megtörténhet. Ennek megfelelően a reálisan létező világ véges méretű, és végtelen volta a sokszoros ismétlődés miatt csak látszat. Ilyen megoldás léte nincs ellentétben az Einstein féle téregyenletekkel. Abból tehát, hogy a galaxisok eloszlásában látható periodikusan ismétlődő csúcsok milyen messze vannak egymástól, meg lehet becsülni ennek a reálisan létező, de a megfigyelésekkel megsokszorozott cellának a méretét.

Az alapvető feltevés a kvazárok esetében is az volt, hogy a periodicitás valójában a távolságban történik, és a vöröseltolódás tulajdonképpen ezt csak tükrözi. Kis kozmológiai távolságokon a vöröseltolódás, és a távolság ugyan arányos egymással, de nagyobb távolságok esetén már ez nem igaz, és a két mennyiség közötti összefüggés konkrét függvényalakja az aktuálisan érvényes kozmológiai modelltől függ.

A homogén, izotróp modellek két alapvető mennyisége a Kozmosz tágulásának üteme (a Hubble állandó, $H_0$), illetve a tágulás ütemének időbeli változása, a gyorsulás paraméter., $q_0$,

Broadhurst és munkatársai úgy találták, hogy az észlelt periodicitás a vizsgált galaxisok eloszlásában a $q_0$=+0.5 értéknél valósul meg a legtisztábban[9]. Amennyiben nem tételezzük fel a $\Lambda$ kozmológiai állandó létét, ez a gyorsulási paraméter annak az Euklideszi határesetnek felel meg ami a pozitív, illetve a negatív görbületű világmodellek között van.

A vizsgált galaxisok eloszlásában kapott periodicitást nagyobb pontossággal reprodukálhatjuk Paál György és munkatársai szerint, ha megengedjük a $q_0$ gyorsulási paraméter negatív értékét is. $q_0$=-0,5 értékkel ugyanis az észlelt adatokhoz sokkal jobb illeszkedést lehet elérni. A negatív értékhez azonban szükség van a pozitív értékű $\Lambda$ kozmológiai állandó feltételezésére. Értékére számszerű becslést is adtak. A számszerű becsléshez szükség volt a tér görbületének a becslésére. Minthogy feltételezték, hogy az észlelt periodicitás a tér többszörösen összefüggő voltának a következménye, arra következtetésre jutottak ez az észlelt módon csak a görbületlen Euklideszi esetben valósulhat meg.

Az Euklideszi eset akkor valósul meg, ha a térben levő összes anyag együttes sűrűsége egy kritikus értékkel egyenlő. A tényleges anyagsűrűségnek ehhez a kritikus sűrűséghez mért arányát $\Omega$-val jelöljük. Nagy kérdés azonban, hogy a kozmológiai állandó mindössze egy matematikai segédeszköz, amivel Einstein téregyenletét ki kell egészíteni, hogy a megfigyelésekkel összhangban levő megoldást nyerjünk, vagy valódi, egyelőre ismeretlen természetű anyagi létező. Az utóbbi esetet feltételezve beszélhetünk a „sötét energia" létéről.

A $\Lambda$ kozmológia állandót anyagi valóságként értelmezve járulékát hozzá kell számítanunk a teljes anyagsűrűséghez. Ennek megfelelően Euklideszi esetben $\Omega=\Omega_\Lambda+\Omega_M=1$. Itt $\Omega_\Lambda$ a sötét energia, míg $\Omega_M$ az összes egyéb anyag arányát jelenti a teljes sűrűséghez. A gyorsulási paraméter $q_0$=-0,5 értéke Euklideszi esetben $\Omega_\Lambda=0,667$, illetve $\Omega_M=0,333$ értékét jelenti. Ezeket az értékeket nézve most tessék mély lélegzetet venni: Az Európai Űrügynökség (ESA) által felbocsátott, 2009-2013 között működő Planck mesterséges hold mérése szerint $\Omega_\Lambda=0,686$! Vagyis a Paál György és munkatársai által 1992-ben becsült érték mindössze 3%-al tér el a jelenleg leginkább elfogadott értéktől!.

Már korábban említettem, hogy a többszörösen összefüggő esetben a végtelen kiterjedés csak látszat, és a valóban létező világ csak egy elegendően nagy cella. Ennek a cellának a méretét Paál György és munkatársai *280/h Mpc* nagyságúra becsülték[10]

---

[9] A pozitív érték azt mutatja, hogy a Világegyetem tágulásának üteme fokozatosan csökken. Tehát lassulást jelent.

[10] Definíció szerint $h=H_0/100$, ahol $H_0$ a Hubble állandó.

**Utószó**

A kozmológiában a 20: százat második évtizedében elkezdődött forradalmi változás első lépése az volt, hogy a cefeidák segítségével a csillagászat kilépett az extragalaxisok világába. Ugyancsak a cefeidáknak köszönhetően állította fel Hubble a róla elnevezett relációt, ami az addig elért kozmikus távolságokat több nagyságrenddel megnövelte[11].

Amennyiben elfogadjuk, hogy az extragalaxisok színképvonalaiban mért vöröseltolódás a távolodásból adódó Doppler-effektus eredménye, a Hubble-effektussal megbecsülhetjük távolságukat. A Hubble-effektusból becsült távolságnak két nagy fogyatékossága van : a belőle számított távolság függ a választott kozmológiai modelltől, illetve, ha a kozmikus anyag mozgásában a távolodásból adódó sebességgel összemérhető egyéb mozgás is van, ennek hatásától a méréseket nem tudjuk elvonatkoztatni.

Fontos tehát egyéb, úgynevezett „szabvány gyertyák" (standard candels) ismerete, amelyek nagy kozmikus távolságból is megfigyelhetők, de távolságuk meghatározásához nincs szükség a Hubble effektusra. Kiderült, hogy ilyen objektumok például az Ia típusú szupernóvák[12], amelyeknél a megfigyelt fénygörbéből következtetni lehet abszolút fényességükre, majd ezt a megfigyelttel összehasonlítva megkaphatjuk a távolságot, a kozmológiai modelltől függetlenül.

Miután kiderült, hogy ezekből a szupernóvából szabvány gyertyák nyerhetők, nagy erőkkel megindult a munka a kozmológiai statisztikus vizsgálatok számára elegendő számú objektum összegyűjtésére.

Paál György 1992-ben bekövetkezett halála után az Ia típusú szupernóvákkal kapcsolatos kutatások drámai erővel felgyorsultak. A kutatások eredményeként a 90-es évek végére kiderült, hogy a Világegyetem gyorsulva tágul. A gyorsulva táguló Világegyetem egyúttal a kozmológiai állandó létének az elismerését is jelentette. Ennek a vizsgálatokból kapott mértéke a hibahatáron belül megegyezik azzal az érékkel, amit Paál György és munkatársai 1992-ben, tehát évekkel azt megelőzően megjósoltak (10. ábra). A felfedezésért Saul Perlmutter, Adam Riess, és Brian Smith 2011-ben elnyerte a fizikai Nobel díjat.

A jelen sorok írója 27 éven át volt kollégája Paál Györgynek a svábhegyi csillagászati intézetben. Sokat beszélgetett vele azokról a problémákról, amin éppen aktuálisan dolgozott. Így sok mindenről a szerzőtől még jóval az eredmények publikálása előtt értesült. Ez a személyes kapcsolat is indított arra, hogy a magam szerény eszközeivel hozzájáruljak ahhoz, hogy Pál György a tudománytörténetben az őt megillető helyet foglalja el.

---

[11] A cefeidák segítségével eddig elért legnagyobb távolság 10 millió parszek körül van, míg a vöröseltolódás mérésével olyan objektumokat is elérhetünk, ahol ez több milliárd parszek is lehet.

[12] William Fowler (fizikai Nóbel díj 1983) és Fred Hoyle elképzelése szerint ilyen szupernóvák egy vörös óriásból és fehér törpéből álló szoros rendszerben jönnek létre. A kettős rendszerben a fehér törpe fokozatosan anyagot nyer a Vörös óriástól, majd egy bizonyos kritikus tömeg (Chandrasekhar határ) elérése után instabillá válik és felrobban.

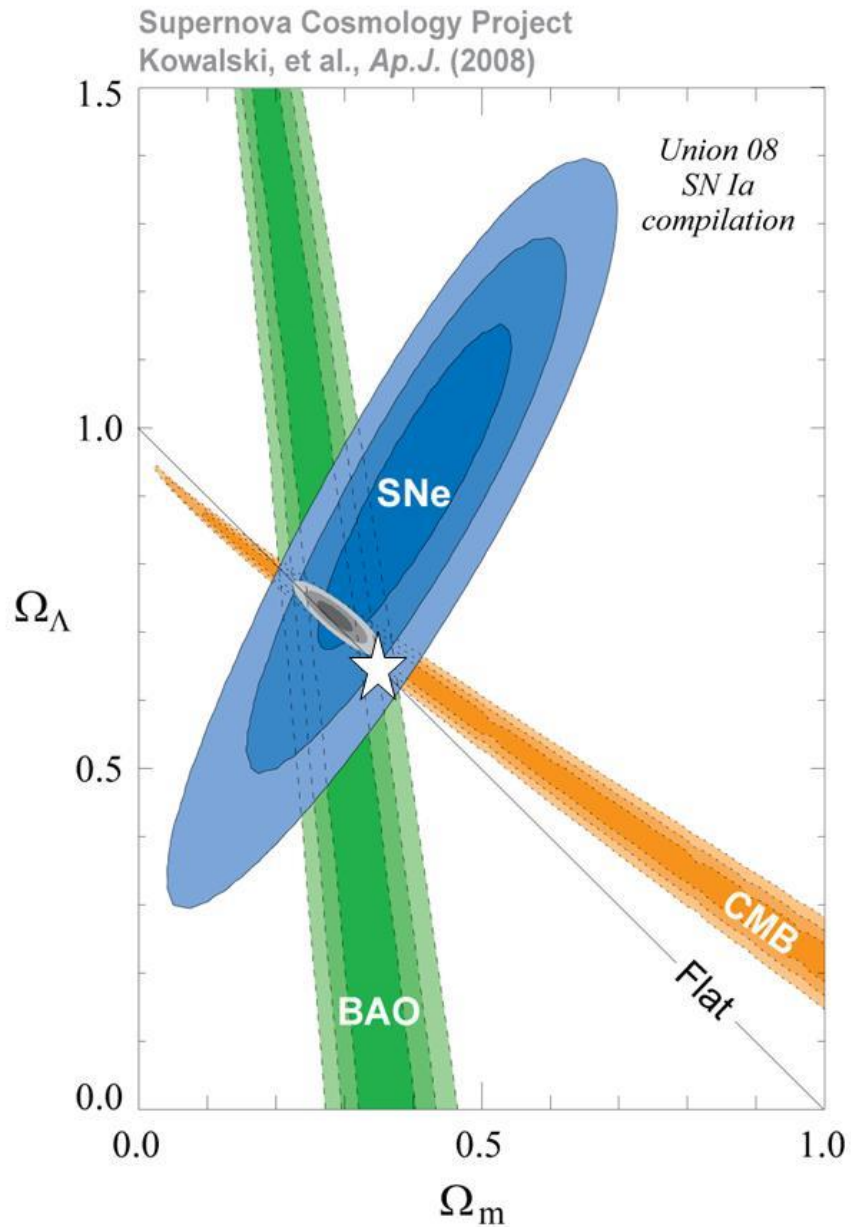

**10. ábra:** A sötét energia, illetve a sötét anyag aránya a sík, euklideszi megoldáshoz szükséges anyagsűrűséghez a szupernóvák (SNe), a kozmikus háttérsugárzás (CMB), illetve a kezdeti sűrűségingadozások (BAO) alapján. A beszínezett területek azokat az értékeket mutatják, ahol a mérési pontosság alapján az értékek lehetnek. A fehér csillag a Paál György és munkatársai által 1992-ben kapott értéket jelöli.